\documentclass[preprint]{aastex}

\def\deg{$^\circ$}

\def\mic{{$\mu$m}}

\def\h2o{H$_2$O}

\def\teff{$T_{\rm eff}$}

\def\aple{$\mathrel{\hbox{\rlap{\hbox{\lower4pt\hbox{$\sim$}}}\hbox{$<$}}}$}

\def\apge{$\mathrel{\hbox{\rlap{\hbox{\lower4pt\hbox{$\sim$}}}\hbox{$>$}}}$}

\begin{document}

\title{The Stellar Content of Obscured Galactic Giant H~II Regions III.: 
W31}

\author{R. D. Blum\altaffilmark{1}} \affil{Cerro Tololo Interamerican
Observatory, Casilla 603, La Serena, Chile\\ rblum@noao.edu}

\author{A. Damineli\altaffilmark{1}}

\affil{ IAG-USP, Av. Miguel Stefano 4200, 04301-904, S\~{a}o Paulo, 
Brazil \\damineli@iagusp.usp.br}

\author{P. S. Conti} \affil{JILA, University of Colorado\\Campus Box
440, Boulder, CO, 80309\\pconti@jila.colorado.edu}

\altaffiltext{1}{Visiting Astronomer, Cerro Tololo Interamerican
Observatory, National Optical Astronomy Observatories, which is
operated by Associated Universities for Research in Astronomy, Inc.,
under cooperative agreement with the National Science Foundation}

\begin{abstract}

We present near infrared ($J$, $H$, and $K$) photometry and moderate
resolution ($\lambda/\Delta\lambda =$ 3000) $K-$band spectroscopy of
the embedded stellar cluster in the giant H~II region W31. Four of the
brightest five cluster members are early O--type stars based on their
spectra. We derive a spectro--photometric distance for W31 of 3.4
$\pm$ 0.3 kpc using these new spectral types and infrared photometry.
The brightest cluster source at $K$ is a red object which lies in the
region of the $J - H$ vs. $H - K$ color--color plot inhabited by stars
with excess emission in the $K-$band. This point source has an $H$
plus $K-$band spectrum which shows no photospheric features, which we
interpret as being the result of veiling by local dust
emission. Strong Brackett series emission and permitted \ion{Fe}{2}
emission are detected in this source; the latter feature is suggestive
of a dense inflow or outflow. The near infrared position of this red source is
consistent with the position of a 5 GHz thermal radio source seen in
previous high angular resolution VLA images.
We also identify several other $K-$band
sources containing excess emission with compact radio sources.  These
objects may represent stars in the W31 cluster still embedded in their
birth cocoons.
\end{abstract}


\section{INTRODUCTION}

\citet[hereafter Paper~I]{bdc99} and \citet[hereafter Paper~II]{bcd00}
presented near infrared imaging and spectroscopic observations of two
optically obscured Galactic giant H~II regions, W43 and W42,
respectively. These observations revealed the massive star clusters at
the center of the H~II regions which had been previously discovered
and studied at longer wavelengths. Our earlier work builds off the
original success by \citet{hhc97} in elucidating the massive stellar
content in M17 through detailed spectral classification of O--type
stars in the $K-$band \citep{hcr96}. The combination of imaging and
spectroscopy at near infrared wavelengths is proving to be a powerful
tool in exploring the youngest stages of stellar evolution and the
birth environments of massive stars. The chief advantage of the near
infrared is that it is long enough in wavelength to lessen (by upto a
factor of ten in {\it magnitudes}) the effect of interstellar
extinction over optical wavelengths and still short enough in
wavelength to potentially probe the stellar photospheric features of
massive stars. Thus, the near infrared is ideal to explore stellar
nurseries to identify and elucidate the nature of their massive stars.

In the present paper, we continue our investigation of the high end of
the Galactic stellar mass function and search for massive young
stellar objects (YSOs) with observations toward W31. W31 is a well
known star forming region which has been observed at radio, far
infrared (FIR), and mid infrared (MIR) wavelengths. The W31 complex
consists of three H~II regions, G10.2--0.3, G10.3--0.1, and G10.6-0.4
\citep{w72}. The first, G10.2--0.3 is a giant H~II region (GHII) which
we define (following the suggestion of Dr. Robert Kennicutt, private
communication) as one which produces at least $10^{50}$ Lyman
continuum ($=$ Lyc) photons per second. This is about ten times the
luminosity of the Orion nebula and roughly the number emitted from the
hottest {\it single} O3-type star.  As these stars are not found in
isolation, there is an implication that a ``giant'' H~II region
contains some minimum of {\it multiple} O-type stars. The embedded
stellar cluster in G10.2--0.3 (which we refer to as W31) is the
subject of this paper. Recent FIR and radio observations as well as a
summary of earlier long wavelength work are presented by
\citet{geal89}.

W31 is located at $l, b =$ 10.2\deg, $-0.3$\deg \ in the first
Galactic quadrant. Its distance is not well known ($>$ 4.1 $\pm$ 0.9
kpc) owing to the classic radio recombination line velocity degeneracy
and a non-definitive molecular absorption velocity \citep{w72}. We
have modified the \citet{w72} lower limit by taking the
sun--to--Galactic center distance as $R_{\rm\circ}$ $=$ 8 kpc
\citep{r93}. We discuss the distance to W31 in \S4 in light of our new
spectro--photometric observations.  \citet{sbm78} put the Lyman
continuum output of W31 at 2$\times$10$^{50}$ s$^{-1}$ (for a distance
of 4.1 kpc, corrected to $R_{\rm\circ}$ $=$ 8 kpc).

\section{OBSERVATIONS AND DATA REDUCTION}

$J$ ($\lambda \approx 1.28$ \mic, $\Delta\lambda \approx 0.3$ \mic),
$H$ ($\lambda \approx 1.63$ \mic, $\Delta\lambda \approx 0.3$ \mic),
and $K$ ($\lambda \approx 2.19$ \mic, $\Delta\lambda \approx 0.4$
\mic) images of W31 were obtained on the nights of 04 May 1999 and 22
May 2000 with the f/14 tip--tilt system on the Cerro Tololo
Interamerican Observatory (CTIO) 4m Blanco telescope using the
facility infrared imager OSIRIS.
\footnote{OSIRIS is a collaborative project between the Ohio State
University and CTIO. OSIRIS was developed through NSF grants AST
9016112 and AST 9218449.}  Spectroscopic data for bright stars in W31
were obtained on the nights of 19, 21, and 22 May 2000 using the f/14
tip--tilt system at the Blanco telescope with OSIRIS. OSIRIS is
described in the instrument manual found on the CTIO web pages
(www.ctio.noao.edu); see also \citet[]{dabfo93}. The
tip-tilt system is described by \citet[]{pe98}. The tip--tilt system
uses three piezo--electric actuators to move the secondary mirror at
high frequency in a computer controlled feed--back loop which corrects
the natural image centroid motion. OSIRIS employs 0.16$''$ pixels in
the mode used for all data described in this paper.

All basic data reduction was accomplished using IRAF\footnote{IRAF is
distributed by the National Optical Astronomy Observatories.}. Each
image/spectrum was flat--fielded using dome flats and then sky
subtracted using a median combined image of five to six frames. For
W31 itself, independent sky frames were obtained 500 arcseconds south
of the cluster. Standard stars used the median combination of the data
for sky.

\subsection{Images}

The OSIRIS 1999 May images were obtained under variable photometric
conditions with some thin cirrus and in $\sim$ 0.5$''$ to 0.6$''$ FWHM
seeing (with the tip--tilt correction). Total exposure times on source
were 504~s, 281~s, and 181~s at $J$, $H$, and $K$, respectively.  The
individual $J$, $H$, and $K$ frames were shifted and combined
(Figure~\ref{w31a}, \ref{w31b}, \ref{w31c}), and these combined frames
have point sources with FWHM of $\sim$ 0.51$''$, 0.54$''$, and
0.56$''$ at $J$, $H$, and $K$, respectively. DoPHOT \citep[]{sms93}
photometry was performed on the combined images. The flux calibration
was accomplished using standard star 9172 from \citet[]{peal98} which
is on the Las Campanas Observatory photometric system (LCO) in
combination with short $J$, $H$, and $K$ images on W31 on 22 May 2000
under photometric conditions.  The LCO standards are essentially on
the CIT/CTIO system \citep[]{efmn82}, though color transformations
exist between the two systems for redder stars.  No transformation
exists between OSIRIS and either CTIO/CIT or LCO; the impact of this
lack of transformation is further discussed below in \S3.4.

The standard observations were made immediately after the W31 data
were obtained and within 0.1 airmass of the airmass for W31. No
corrections were applied for these small differences in
airmass. Aperture corrections using 20 pixel radius apertures were
used to put the instrumental magnitudes on a flux scale using between
five and eight relatively uncrowded stars on the W31 images.

Uncertainties for the final $J$, $H$, and $K$ magnitudes include the
formal DoPHOT error added in quadrature to the published error of the
photometric standards and to the uncertainty of the aperture
corrections (between the OSIRIS 1999 and 2000 data and the OSIRIS 2000
data and the standard star).  The sum in quadrature of the aperture
correction and standard star uncertainties is $\pm$ 0.027, $\pm$
0.018, $\pm$ 0.021 mag in $J$, $H$, and $K$, respectively. The DoPHOT
errors ranged from approximately $\pm$ 0.01 mag to an arbitrary
cut--off of 0.2 mag (stars with larger errors were excluded from
further analysis).

The flat--field illumination for the 1999 May images was not uniform.
A smooth gradient with full range of about 10$\%$ was
present. Corrections to the final photometry were made based on
observations of a standard star taken over a 49 position grid covering
the array.

\subsection{Spectra}

The $K-$band spectra of six of the brightest stars in the center of
W31 were obtained with a 0.48$''$ wide slit (oriented EW) in \aple
0.5$''$ to 7$''$ FWHM seeing and divided by the spectrum of HR~6953
(B9V), HR~6962 (A2V), or HR~7183 (B8V) to remove telluric absorption
features. Total on source integration times were between 10 and 21
minutes for each W31 object and about 100 seconds for each telluric
standard.  Br$\gamma$ absorption in the HR stars was removed by eye by
drawing a line across it between two continuum points. One dimensional
spectra were obtained by extracting and summing the flux in $\pm$ 2
pixel apertures (0.64$''$ wide).  The extractions include background
subtraction from apertures centered \aple 1.0$''$ on either side of
the object.

An $H-$band spectrum was obtained of the brightest cluster object,
W31~\#1. The slit width, telluric correction, seeing, and extraction
were similar to those for the $K-$band spectra. HR~6962 (A2V) was used
to correct for telluric absorption. The intrinsic
Br absorption in HR~6962 was removed by
making smooth fits to the line profiles, and then dividing the original
spectrum by the fit line spectrum. Regions outside the line profiles 
were set to unity in the fit line spectrum.
The total on source integration time was
14 minutes.

The wavelength calibration was accomplished by measuring the positions
of bright OH$^-$ lines from the $H-$band sky spectrum
\citep[]{oo92}. Lines in the W31 spectra are identified by their
relative differences between one and another. The measured dispersion
is 0.0002762 \mic \ pix$^{-1}$ in $H$ and 0.0003683 \mic \ pix$^{-1}$
in $K$. The spectral resolution at 1.65 and 2.2 \mic \ is
$\lambda/\Delta\lambda \approx$ 3000.

\section{RESULTS}

The $J$, $H$, and $K-$band images shown in
Figure~\ref{w31a}--\ref{w31c} have been combined into a false color
image (Figure~\ref{w313c}) with colors red, green, and blue
representing $K$, $H$, and $J$, respectively. Figure~\ref{w313c}
reveals a rich cluster of newly formed stars at the heart of W31.
Some of these (see below) are shown to be massive O stars from their
$K-$band spectra. Spectroscopic and photometric evidence is also given
below which indicates that young stellar objects (YSO) are present;
thus the W31 cluster must be very young and its stars still undergoing
their birth pangs.

The bright object (source ``A'') near the center of Figure~\ref{w313c}
is a foreground star.  Our unpublished $K-$band spectrum and its
position in the color--color plot (see below) suggest it is an M dwarf
star. We have identified this object (and several others) on the
Digital Sky Survey
\footnote{
Based on photographic data obtained using The UK Schmidt Telescope.     
The UK Schmidt Telescope was operated by the Royal Observatory          
Edinburgh, with funding from the UK Science and Engineering Research    
Council, until 1988 June, and thereafter by the Anglo-Australian        
Observatory.  Original plate material is copyright (c) the Royal        
Observatory Edinburgh and the Anglo-Australian Observatory.  The        
plates were processed into the present compressed digital form with     
their permission. The Digitized Sky Survey was produced at the Space   
Telescope Science Institute under US Government grant NAG W-2166.} 
red image.  The USNO~A2 catalog gives the coordinates of this
object as $\alpha$~(2000) = 18$^{\rm h}~09^{\rm m}~26.32^{\rm m}$,
$\delta~(2000) = -20$\deg$~19'~13.7''$.

\subsection{Color Magnitude and Color--Color Diagrams}

The $H - K$ vs. $K$ color--magnitude diagram (CMD) and $J - H$ vs. $H
- K$ color color plot are shown in Figures~\ref{cmd} and \ref{cc}. A
foreground sequence is seen in Figure~\ref{cmd} at $H - K$ $\sim$ 0.5
mag. To the red and lying at $H - K$ between about 1.0 and 2.0 is the
W31 main sequence. Some stars for which spectra are presented below are
labeled. The relative colors of the reddest stars can be used
to distinguish two sequences in Figure~\ref{cc}. There is a well
populated sequence beginning near $J - H$ vs. $H - K$ $=$ 0, 0 which
is taken to be the reddened main sequence. A second sequence lying to
the red in $H - K$ is populated by sources with a clear indication of
excess emission and much larger extinction. We argue below that these
are young stellar objects (YSOs; e.g. \citet{la92}; see also the
discussion in Paper~II), though the precise evolutionary state is
difficult to pin down from near infrared observations alone. One of
these objects, W31~\#1, was observed spectroscopically and is
discussed further in \S3.2 and \S4.

\subsection{Spectra}

Spectra for each of four hot stars in the W31 cluster are presented in
Figure~\ref{spec}. We name these objects W31~\#2, \#3, \#4, and \#5 as
they are the second to fourth brightest sources at $K$ in
Figure~\ref{cmd}, excluding foreground sources (defined as objects
with $H - K$ $<$ 0.6 mag). The $H$ plus $K$ spectrum of W31~\#1 (the
brightest cluster source) is shown in Figure~\ref{hspec}. All the
spectra have been ratioed by an A--type star as described in \S2 to
correct for telluric absorption; the spectra in Figure~\ref{spec} have
been further normalized by a low order fit to the continuum. In the
case of W31~\#1, the spectrum has been put on a flux scale by
multiplication by an appropriate blackbody (10000 K). The $H$ and
$K-$band segments are normalized based on the photometry of
Figure~\ref{cmd}.  The spectrum in Figure~\ref{hspec} is not corrected
for extinction. The upturn in the continuum at the end of the $H-$band
is probably due to vignetting in the OSIRIS camera optics; 
the continuum slope 
in this wavelength region is not reliable though
the line profiles should not be substaintially afected relative to the local
continuum. An analogous downturn may be present at the very end of the
$K-$band.

\subsubsection{O Stars}

The spectra of sources W31~\#2--5 may be compared to the $K-$band
spectroscopic standards presented by \citet[]{hcr96}. The features of
greatest import for classification are (vacuum wavelengths) the
\ion{C}{4} triplet at 2.0705, 2.0796, 2.0842 \mic \ (emission), the
\ion{N}{3} complex at 2.116 \mic \ (emission), and \ion{He}{2} at
2.1891 \mic \ (absorption). The 2.0842 \mic \ line of \ion{C}{4} is
typically weak and seen only in very high signal to noise spectra
\citep[]{hcr96}. The presence of \ion{N}{3} and \ion{He}{2} leaves no
doubt that these are O--type stars, and the \ion{C}{4} emission firmly
places each in the kO5--O6 subclass. This class is predominantly made
up of stars with optical MK spectral types of O5 to O6, but several O4
stars presented by \citet[]{hcr96} also show \ion{C}{4} emission.
None of the spectra in Figure~\ref{spec} show \ion{He}{1} 2.12 \mic \
absorption which is present in later type O7-9 stars.

Generally strong absorption in Br$\gamma$ is expected for kO5--6 dwarf
and giant stars and weak absorption or emission for super--giants, so
the absorption wings seen in Figure~\ref{spec} suggests the stars are
not highly evolved. However, there is also residual emission seen in
W31~\#2, 4, and 5.  The emission appears superposed on a broader
absorption profile and might otherwise be suggestive of a
circumstellar origin (e.g. in a disk, though there is no evidence of
excess emission for these stars in Figure~\ref{cc}) were it not for
the following.  There are two problems with Br$\gamma$ which make it
difficult to interpret and thus render any observations
speculative. The first is the fact that Br$\gamma$ has to be corrected
in the telluric standard (\S2), and this is only accurate to five or
ten percent for the method used here. Since the correction is for
absorption in the standard and is typically under corrected (the core
of the line is straight forward to identify, but the wings and
continuum more difficult), the tendency is to reduce the intrinsic
absorption in the object. It is also possible that any residual
absorption in the telluric standard would have a different profile and
position than the object. The second problem is that the nebular
emission, which is subtracted by using background apertures, can vary
over small scales. Even though the background is an average from
either side of the object and in close spatial proximity to it, there
is no guarantee that it is uniform or linearly varying across the
source. It is interesting to note that we find no residual emission 
from \ion{He}{1} 2.06 \mic. 
Though Br$\gamma$ is often seen in emission in O supergiants
\citep{hcr96}, the character of the emission is not like that shown in
Figure~\ref{spec} with absorption superposed upon emission.  In
summary, we can not tell if the residual emission in the O--stars is
real (i.e. circumstellar) or an artifact. Higher spectral and spatial
resolution observations will be required to make further progress.

\subsubsection {W31 \#1}

The spectrum of W31~\#1 is shown in Figure~\ref{hspec}.  The spectrum
rises more steeply to the red relative to the unratioed spectra of the
O stars discussed in \S3.2.1 (only the continuum divided O star
spectra are shown here). This suggests a larger line-of-sight
extinction and/or excess emission at redward wavelengths than for the
O stars. There are no discernible photospheric absorption features
present in either the $H$ or $K$ bands, though the signal-to-noise is
relatively high, \apge 70, in the fainter $H-$band. Besides the lines
discussed above for the $K-$band, lines of \ion{He}{1} (1.700 \mic)
and \ion{He}{2} (1.692 \mic \ and 1.572 \mic) are expected for
luminous hot stars \citep[]{brcfs97, hrl98}. \ion{Fe}{2} emission is
identified at (1.6878 \mic) and perhaps also at 1.7419 \mic; see
Figure~\ref{feii}. 
There is weak \ion{He}{1} 2.06 \mic \ emission seen in Figure~\ref{hspec};
however \ion{He}{1} 1.700 \mic \ is absent from the $H-$band spectrum
of W31~\#1.  These facts suggest that the line producing region seen
in W31~\#1 is not very hot.

\subsection{Mid and Far Infrared Sources}

Our position for the stellar cluster places it near ($\sim$ 38$''$NW)
the strong far infrared peak S9 presented by \citet{geal89} and to the
SE of the even more luminous far infrared source S7. The far infrared
beam was several arcminutes in diameter and the positional uncertainty
quoted is $<$ 1$'$. The position given by \citet{fpa77} for the mid
infrared source toward G10.2--0.3 (29$''$ aperture diameter) is
approximately 38$''$ east of the near infrared cluster center;
\cite{fpa77} quote coordinates good to about 20$''$. S9 and the mid
infrared source are thus possibly coincident with each other and also
with the stellar cluster. However, we believe the positional offsets
should be taken at face value. It is more likely that the stellar
cluster has cleared away much of the gas and dust in its immediate
vicinity and that the hottest stars are exciting the material
surrounding this cavity. This situation is similar to that for W43
(Paper I). The dust distribution is apparently non--uniform and the
exciting cluster gives rise to the strong far infrared sources S9 and
S7 to the SE and NW of the cluster, respectively.

\subsection{Extinction}

The results of the preceding sections may be used to estimate the
total line--of--sight extinction to individual sources in the W31
cluster. However, as discussed in \S~2, the OSIRIS colors are not on a
standard system. The largest effect is expected to be in the $J$
filter \citep[Paper II]{peal98, eegb97} such that the observed $J - H$
and $J - K$ colors are more red than CIT/CTIO. There is no evidence
for a significant color term in $H - K$ \citep{peal98}.

An estimate of the extent to which color--terms are important can be
made by comparing the W31 colors to the {\it dashed} reddening lines
in Figure~\ref{cc}. The intrinsic colors for O stars and red giants
are taken from \citet{k83} and \citet{fpam78}, respectively. The
former colors have been put on the CIT/CTIO system (Paper II) by
making minor (\aple 1$\%$) corrections; the latter colors are on the
CIT/CTIO system. The reddening lines themselves are derived using the
interstellar extinction law of \citet{m90} which is itself based on
average relations from multiple photometric systems. The range of
values for the near infrared extinction laws considered by Mathis
result in a difference of about 0.15 mag for the lines shown in
Figure~\ref{cc} at $H - K =$ 2.0 mag. There is a trend for the OSIRIS
colors to be too red in $J~-~H$ evident in Figure~\ref{cc} which is in
agreement with expectations as noted above. If we were to apply the
same transformation to the $J-$band colors as for the CTIO facility IR
imager CIRIM \citep{eegb97} which has similar filters and detector as
OSIRIS, then the effect would be for the reddened main sequence of the
cluster to more closely match the lower reddening line in
Figure~\ref{cc} and for the stars with excess to lie even further
below the line. The correction would be about $-$0.2 mag in $J - H$
for the red stars like W31~\#1 in Figure~\ref{cc}. This suggests that
the magnitude of the undetermined color transformation to put OSIRIS
magnitudes on the CIT/CTIO system is consistent with a negligible
correction for $H - K$.

We will therefore assume that the $H - K$ colors of OSIRIS are the
same as for the CIT/CTIO system and compute estimates of the
extinction based on $H - K$ alone and the Mathis interstellar
extinction law. Table~\ref{ostar} lists the photometry and derived
$A_K$ for the four O-type stars presented in this paper. The mean
extinction toward these four objects is $A_K =$ 1.71 mag $\pm$ 0.12
mag where the uncertainty is the standard deviation in the mean. We
add to this value (in quadrature) 0.15 mag due to the expected range
in the power--law exponent of the \citet{m90} extinction law ($\pm$
0.1 in the exponent).

\section{DISCUSSION}

W31 hosts a complex mixture of newly formed stars, gas, and dust as
Figure~\ref{w313c} attests. The spectra presented in \S3 indicate that
some of these newly formed stars are quite massive and also that
massive stars are still forming, or at least have not yet shed their
natal material and revealed their photospheres. In the following
sections we discuss these spectra and the stellar and young stellar
content in more detail, establishing the basic properties of the W31
cluster.

\subsection{Distance}

As mentioned in \S1, the distance (and hence luminosity) of W31 has
not been well known. \citet{w72} discusses the radio recombination
line velocity for W31. The Galactic rotation model is degenerate for
this case, but the absorption due to H$_2$CO in intervening clouds is
significantly higher than the H recombination line velocity indicating
the absorption is not local and that the H~II region is not at the
near distance. Interestingly, \citet{gg76} placed W31 at the {\it far}
distance of 18.7 kpc; it is the most distant GHII region in their
famous plot (their Figure 11) of Galactic structure, well beyond the
Galactic center. On the other hand, the H$_2$CO absorption does not
extend to the maximum velocity on the line--of--sight, so only a
minimum, intermediate distance, $D >$ 4.1 $\pm$ 0.7 kpc can be
estimated from this method. Even this limit is quite uncertain as the
H$_2$CO may be affected by non--circular velocities.

We can now constrain the distance to W31 using the spectroscopic and
photometric results of \S3. We compute distances assuming the O stars
shown in Figure~\ref{spec} are ZAMS or dwarf luminosity class
(i.e. hydrogen burning). The results of the following sections firmly
demonstrate that W31 is a very young cluster; hence, we will not
consider the O stars to be evolved giants for which ages would be
between about 2.5 and 5 Myr \citep{ssmm92}. The near infrared spectra
of the O--stars are also not consistent with those of supergiants
(\S3.2.1). The spectral type in each case is assumed to be O5.5V,
consistent with the infrared type assigned above, and only the
luminosity is varied. For the ZAMS case, the $M_K$ is taken from Paper
II. For the dwarf case, the distance is determined using the $M_V$
given by \citet{vgs96} and $V - K$ from \citet{k83}. In all cases the
distance estimate uses the photometry and $A_K$ given in
Table~\ref{ostar}. The distance estimates are shown in
Table~\ref{ostar} and result in mean distances of 3.1 $\pm$ 0.3 kpc
and 3.7 $\pm$ 0.4 kpc for the ZAMS and dwarf cases, respectively. The
uncertainty quoted for the mean distance is the standard deviation in
the mean of the individual distances added in quadrature to the
uncertainty in $A_K$ ($\sim$ 250--350 pc).

The \citet{w72} distance is a lower limit. Comparing this limit with
the spectro--photometric estimates above, the radio limit is
consistent (within one $\sigma$) with the estimates using the dwarf or
ZAMS luminosity. As mentioned above, we argue below that W31 is very
young based on the presence of YSOs. Thus, we do not expect the
O--type stars to be highly evolved and we can firmly constrain the
distance to be within 3.1 and 3.7 kpc with the expectation that it may
be at or near the ZAMS distance. We will adopt a mean distance of 3.4
kpc. Then, the region shown in Figure~\ref{w31a} is about 1.7 pc
across, and that of the central cluster 0.6 pc.

\subsection{Luminosity and Age}

We can estimate the age of the O stars by placing them in the
Hertzprung--Russell diagram (HRD). The average O star luminosity was
set in the previous section by adopting the intrinsic luminosities
consistent with the observed spectral types (the average of
the ZAMS and dwarf cases) in order to derive the distance to the
cluster. To place the individual stars in the HRD, we use this average
distance (3.4 kpc), the bolometric correction vs. spectral type
relationship given by \citet{vgs96}, the photometry of
Table~\ref{ostar}, and the $V - K$ for O--type stars presented by
\citet{k83}.  The result is shown in Figure~\ref{hrd}. The uncertainty
in $M_{\rm bol}$ is due to an assumed distance uncertainty of $\pm$
0.3 kpc and the uncertainty in $A_K$ and $K$ as given in
Table~\ref{ostar}. The \teff \ uncertainty is taken as the range in
\teff \ for the O5--O6 spectral types as tabulated by
\citet{vgs96}. The data points are compared to the model isochrones of
\citet{ssmm92}. Figure~\ref{hrd} indicates that the W31 O--type stars
are quite young, less than about one million years with a mean age of
perhaps 500 thousand years.

The Lyman continuum (LyC) output of W31 cited by \citet{sbm78} is
2$\times$10$^{50}$ sec$^{-1}$ (for $R_{\rm\circ} = 8$ kpc). For a
distance of 3.4 kpc to W31, this number becomes 1.4$\times$10$^{50}$
sec$^{-1}$ (14 O star equivalents -- \citet{vgs96}). 
The LyC output for a single O5.5V star,
similar to what we observe, is 2.7$\times$10$^{49}$ sec$^{-1}$. Thus,
five such stars can account for the total ionizing radiation
production estimated from radio measurements. For the distance of 3.4
kpc derived in the previous section, the average $M_{\rm bol}$ of the
O stars is about $-$9.1 mag (Figure~\ref{hrd}). This is about 0.2 mag
fainter than the O5.5V luminosity, so the corresponding LyC value is
also about 20$\%$ lower. If the O stars are truly on the ZAMS, then
the distance, and hence average luminosity, is an additional 20$\%$
lower than the O5.5V value (i.e. 0.4 mag total). There is thus room
for $\sim$ 5$-$9 O5.5 stars; i.e. we have found four hot stars which
account for about 50 to 80 $\%$ of the ionization indicated by radio
emission (which does not account for destruction by dust or loss by
leakage). \cite{geal89} and \citet{whp84} presented 5 GHz maps of the
W31 region. They suggested that many of the sources in their maps
could be embedded young stars. Some of these objects may contribute to
the ionization in W31, though in light of the smaller distance derived
above, none of their sources is particularly luminous (their most
luminous source overlapping with our images would be consistent with a
O9 or B0 star). We further discuss below the \citet{geal89} sources in
conjunction with near infrared excess emission objects presented in
this paper.

\subsection{Star Formation}

In the midst of the complex gas and dust morphology evident in
Figure~\ref{w313c}, we have identified four luminous O--type
stars. Figure~\ref{cmd} indicates there are more which are not yet
spectroscopically confirmed. And more importantly, Figure~\ref{cc}
indicates there are a number of red objects which might be still
dramatically affected by their natal dust clouds. These are the
objects which appear offset to the red from the normal sequence of
stars in Figure~\ref{cc}; see the object labeled \#1 in the Figure. A
clear excess emission (at least in $K$) is indicated by the offset.
This excess is attributed to accretion luminosity in a circumstellar
disk for lower mass YSOs \citep{mch97,la92} or perhaps circumstellar
envelopes in the case of Herbig Ae/Be (HAEBE) stars
\citep{hkc93,psl97}. \citet{la92} identified lower mass objects in
this region of the color--color plot as possible protostars. The
protostars are distinguished from more evolved PMS stars by their
location farther along the reddening line indicating extinction by
more obscuring circumstellar material. Accounting for a foreground
extinction as derived for the O--type stars in Table~\ref{ostar}, then
the YSO candidates in Figure~\ref{cc} would lie near the lower end of
the region identified by \citet{la92} as being populated by
protostars. However, \citet{la92} did not attempt to account for
foreground extinction in their YSO sample.

In Papers I \& II, we identified similar YSO candidates in W43 and
W42. These objects are likely massive stars in the process of shedding
their birth material. Like the YSOs identified in M17 by
\citet{hhc97}, the objects in W43 and W42 may be of somewhat lower
mass (based on their position in the CMD) than the more luminous
spectroscopically identified O stars (perhaps late O--type or early
B--type; see \citet{hhc97}). In M17, the more massive stars showed no
evidence for circumstellar disks, but several later type stars did. An
obvious question is whether the most massive stars formed disks at
all, but the case in M17 is not conclusive because the stars with disk
signatures are spatially segregated from the most massive O--type
stars.

Several possibilities exist. All the massive stars are coeval, but the
most massive either formed with no disks or the active disk phase was
rapid and now beyond detection. Or, perhaps the later types simply
formed later in a sequential mode. It is significant that the
object W31~\#1 is both spatially located in the heart of the cluster
and more luminous at $K$ than any of the O--stars, unlike the
suggestion of a spatial segregation of the stars with disks and
without in M17. 

In any event, the near infrared excess emission alone does not
indicate the physical source of emission (e.g. from a disk or
envelope), but it is a clear indication that this is a YSO. W31~\#1 is
more than 0.5 mag brighter at $K$ than any of the O stars listed in
Table~\ref{ostar} indicating it is a massive YSO.

The spectrum of W31~\#1 shown in Figure~\ref{hspec} shows no sign of
photospheric hot--star features. Therefore, it does not allow us to
determine whether it is a ZAMS late-O or early-B--star which has yet
to clear away its remnant natal material (like a compact H~II region)
or perhaps a younger object just transitioning from an earlier
proto--stellar phase. The Brackett series emission suggests an
internal ionizing source (recall that background apertures, \S2.2,
have been used in an attempt to remove the nebular emission from the
circumstellar emission). The connection between dense, hot molecular
cores which might be massive protostars in a collapse or accretion
phase and ultra compact H~II (UCHII) regions which represent
enshrouded but hydrogen burning objects (albeit very young) is the
subject of intense recent study; see \citet{kcchw00} for a
review. These objects probably form an evolutionary sequence
\citep{kcchw00} and the former are often sought by observations toward
the latter. Figure~\ref{feii} shows a detailed view of the $H-$band
spectrum of W31~\#1. There is both Brackett emission and \ion{Fe}{2}
emission present. Such permitted lines are indicative of dense flows
or winds and require large amounts of UV photons and dense gas (\apge
10$^6$ cm$^{-3}$), for example, in the massive evolved star Eta Carina
\citep{hdje94}. The \ion{Fe}{2} emission is narrow, and may also be
consistent with dense material in a circumstellar disk.

Thus the emission--line spectrum of W31~\#1 is consistent with the
picture that a massive object lies hidden within a large amount of
circumstellar material. This object would be sufficiently massive to
produce a dense outflow/inflow and to produce sufficient ionizing
photons to ionize the flow/wind. Massive protostars may remain
``hidden'' (i.e. they are not detected as UCHII) during the
collapse/accretion phase because the mass infall rate is sufficient to
arrest the development of an H~II region around the ionizing protostar
\citep{w95}. Since we see evidence for ionized circumstellar material
around W31~\#1, this suggests it is not still in such an early
proto--stellar phase. In this case we might expect W31~\#1 to be
associated with one of the radio sources reported by \citet{geal89};
i.e. it would be an UCHII region.  In Figure~\ref{radio} we plot an
overlay of the $K-$band image of Figure~\ref{w31a} with the positions
of the \citet{geal89} radio sources (5 GHz sources from their Table
2). The \citet{geal89} 5 GHz beam was 7.5$''$$\times$4.2$''$ FWHM. As
can be seen in the figure, there is a radio source near the position
of W31~\#1. This is radio source 23 from \citet{geal89}, and it is about
4$''$ from the near infrared position of W31~\#1. Thus, W31~\#1 is a
candidate UCHII region. 

W31~\#1 is not the only YSO candidate in the cluster; in fact, it is
one of five bright objects in the excess region of the color--color
plot (Figure~\ref{cc}). Each of these objects (W31~\#1, 9, 15, 26, 30)
is in the bright end of the $K$ vs. $H - K$ CMD, so is potentially
massive (see Table~\ref{yso}). They all appear to be forming in the
central region of the cluster where massive stars may be
preferentially expected; see, for example, \citet{hh98}. These objects
are indicated in Figure~\ref{w31a} with {\it open circles}. The
position of W31~\#26 is consistent with the 5 GHz radio source 21 of
\citet{geal89}. This is the brightest of the Ghosh et al. sources in
our field (and one of the brightest in their entire list). 

In Figure~\ref{radio}, we plot the positions of the remaining Ghosh et
al. radio sources as well as their radio 
sources 21 and 23 discussed above.  We searched
our near infrared source list for candidate matching objects to each
of the remaining Ghosh sources. We have identified the reddest object
within the radio FWHM as a possible candidate and show the results in
Table~\ref{ghosh}. We have identified a potential radio counterpart in
all cases except for Ghosh radio source 15. Though there is no spectroscopic
information for these sources, we suggest that they may also be UCHII
regions based on their infrared colors and likely conincidence with
the 5 GHz radio sources. All the candidate counterparts identified in
Table~\ref{ghosh} are sufficiently red that they have no $J-$band
detections (except for radio sources 
21 and 23 as discussed above). Since the radio sources are
relatively bright at $K$, the lack of $J-$band detections
suggests they also exhibit excess
emission, though we can't prove it.  Radio source 15 has no obvious
counterpart on the $K-$band image. If the radio source represents an
embedded object, then it is still too buried to be seen at shorter
wavelengths and may represent the most recent, on--going, star birth
in the W31 cluster. The last three candidates in Table~\ref{ghosh} may
be too faint to be associated with the radio sources, as the implied
$K$ magnitudes are inconsistent with an interpretation as early
B--type stars. If so, the true sources may still be buried and
invisible at near infrared wavelengths, for example, as are the ``western''
UCHII regions in W49 -- \citep[]{seal00}.

We have tabulated the Lyman continuum luminosity for each of the
\citet{geal89} sources in our field in Table~\ref{ghosh}. These values
have been corrected for the distance derived above, and so are
somewhat less luminous than reported by \citet{geal89}. We also
tabulate spectral types based on these values using the derived
properties of dwarf OB stars from \cite{vgs96}. Most are about
B0. Could W31~\#1 harbor such an embedded source?
If we assume the excess emission from W31~\#1 is due to reprocessing of the
stellar radiation by a circumstellar disk, we can estimate its intrinsic
stellar luminosity. Using the tabulation of \cite{hsvk92} for reprocessed
radiation from a flat circumstellar disk, we find that $\Delta K$ $=$ 3.71
mag and $\Delta (H-K)$ $=$ $-0.54$ mag. This is the maximum luminosity that such
a reprocessing disk may produce and assumes that the disk is viewed 
along the line of sight (it is seen along an axis perpendicular to the 
plane of the disk)
and extends to the stellar surface (there is no hole in the disk).

Adjusting the observed magnitudes for this putative disk,
we find that the $A_K$ would be
2.9 mag for the \citet{m90} extinction law. This results in an intrinsic $K$ 
magnitude which is 1.71 mag fainter than the O stars in W31, equivalent to a 
B0.5 star (Paper II).
This luminosity or spectral type estimate is consistent with the \citet{geal89}
estimate from the radio continuum observations; see Table~\ref{radio}.
Such an object does not have a very hard ionizing continuum and 
so might also explain the lack of neutral helium emission in 
Figure~\ref{hspec}. Low ratios ($< 10\%$) of \ion{He}{1} to Br$\gamma$ are
expected for such stars \citep{bd99}.
Recall that one of the other YSO candidates, W31~\#26, is associated with
the brightest radio source in our field, radio source 21.
An inclined circumstellar disk would explain why its apparent $K-$band excess
is not as prevalent as that for W31~\#1 (see Figure~\ref{cmd} and
Table~\ref{radio}).

While illustrative, our
model for the production of the near infrared excess in W31~\#1 is not
unique. We have not considered accretion luminosity contributing to the 
$K-$band emission or dust destruction of the ionizing radiation in the
UCHII region, for example.
We plan to make longer wavelength observations in the mid--infrared 
to further explore the nature of this object by expanding the wavelength 
baseline to better constrain the source of its circumstellar emission.
In general, we envision a situation where the
O--stars and most of the YSO candidates are revealed and are ionizing
the nebula in W31. The majority of the Ghosh et al. sources, however, 
may be less evolved, optically thick, UCHII regions.

It is interesting to further speculate on the nature of the star birth
history in W31. If the Ghosh et al. sources are indeed associated with
the near infrared sources, then their radio luminosities
(Table~\ref{ghosh}) indicate they are generally less massive than the
spectroscopically identified O--stars. We then have a situation
similar to what may have occured in M17 \citep{hhc97} where the most
massive stars are revealed first with the later type O and B stars
taking longer to clear away their natal material (or being slower to
form). In the present case, there may also be on--going star birth of
somewhat less massive hot stars (the buried radio source 15 of
\citet{geal89}).  A further similarity to M17 is that the most massive
O stars appear not to show obvious disk signatures, and the implication is
that the active disk phase for the most massive O stars is shorter than for
lower mass O and B stars or that these objects don't form disks at all. 
In the present case we can not
yet firmly place the YSO W31~\#1 on a mass scale,
but its near infrared and radio properties are
consistent with an early B--type star surrounded by circumstellar material.

\section{Summary}

We have presented initial, yet detailed, observations of the stellar
and young stellar content embedded in the giant H~II region W31. Our
$J$, $H$, and $K$ images reveal a dense stellar cluster in W31,
including young stellar objects. $K-$band spectra are presented for
five objects. Four of these are identified as early O--type by
comparison to standard stars. The fifth and brightest cluster member
at $K$
has near infrared colors which are similar to known young stellar
objects.  Strong excess $K-$band emission places it to the red of the
normal stellar sequence in the $J - H$ vs. $H - K$ color--color plot.
We find that the position of this object, W31~\#1, is consistent with
a 5 GHz thermal radio source and suggest it therefore may be an ultra
compact H~II region, though perhaps one with an early B--type exciting
star. It has an $H$ and $K-$band spectrum devoid of
photospheric lines, indicating it is still heavily enshrouded by
circumstellar material. The Brackett lines are in emission.  Its
$H-$band spectrum exhibits permitted \ion{Fe}{2} line emission which
we suggest indicates a dense circumstellar environment, perhaps a
strong inflow, outflow or circumstellar disk. 

We identified four other YSO candidates based on photometry in the $J
- H$ vs. $H - K$ color--color plot. These stars (which may also be
massive as indicated by their $K-$band magnitudes) and W31~\#1 are
found in the heart of the W31 cluster among the O--type stars. We also
identified a thermal radio source with one of these YSO
candidates. Further, for the remaining 5 GHz sources in our field, we
identified potential $K-$band counterparts with offsets less than the
size of the FWHM of the radio observations (\aple 5$''$). We suggest
that these may be ultra compact H~II regions representing the most
recent massive star birth in the cluster.

Spectral types of the newly identified O--stars and the photometry
presented herein firmly constrain the distance to W31 which was
previously uncertain from radio observations.  Based on the $K-$band
spectra and presence of YSOs, we argued that the O--stars can not be
highly evolved stars (i.e. giants or supergiants). We then computed
distances assuming the O--star spectra were consistent with ZAMS or
dwarf luminosity classes by taking the appropriate range of intrinsic
luminosities for O--stars. We find the distance to W31 must be less
than about 3.7 kpc which is consistent with the previous lower limit
from radio studies. Shorter distances (about three kpc) would be
inferred for the O--stars (but still consistent with the radio data
within the uncertainties) if they are precisely on the ZAMS. The
inferred age of the O--stars is about half a million years or less.

PSC appreciates continuing support from the National Science
Foundation. AD thanks PRONEX and FAPESP for support.
We appreciate the comments of an anonymous referee which have helped to improve our paper. We are grateful to S. Strom for useful conversations regarding
young stellar objects.

%REFERENCES

\newpage

%FIGURES

\begin{figure}
\figurenum{1$a$} 
%\plotone{Blum.fig1a.eps} 
\figcaption[]{$K-$band image of
the massive star cluster in W31.  North is up, East to the left, and
the scale is 0.16$''$ pix$^{-1}$ in this $\sim$ 1.8$'$ $\times$ 1.7$'$
image. The star marked ``A'' is a foreground object; see text. Stars
\#1--5 have spectra presented in Figures~\ref{spec} and \ref{hspec}
(see also Table~\ref{ostar}. The white {\it circles} indicate 
candidate young stellar objects; see text and Table~\ref{yso}.  }
\label{w31a}
\end{figure}

\begin{figure}
\figurenum{1$b$}
%\plotone{Blum.fig1b.eps}
\figcaption[]{Same as for Figure~\ref{w31a}, but for the $H-$band.
}
\label{w31b}
\end{figure}

\begin{figure}
\figurenum{1$c$}
%\plotone{Blum.fig1c.eps}
\figcaption[]{Same as for Figure~\ref{w31a}, but for the $J-$band. Note 
that at this wavelength a star cluster is no longer obvious. }
\label{w31c}
\end{figure}

\begin{figure}
\figurenum{2} 
%\plotone{Blum.fig2.eps} 
\figcaption[]{False color image ($K
=$ red, $H =$ green, $J =$ blue) constructed from the images shown in
Figures~\ref{w31a}, \ref{w31b}, \ref{w31c}.  North is up, East to the
left.  }
\label{w313c}
\end{figure}

\begin{figure}
\figurenum{3} 
\plotone{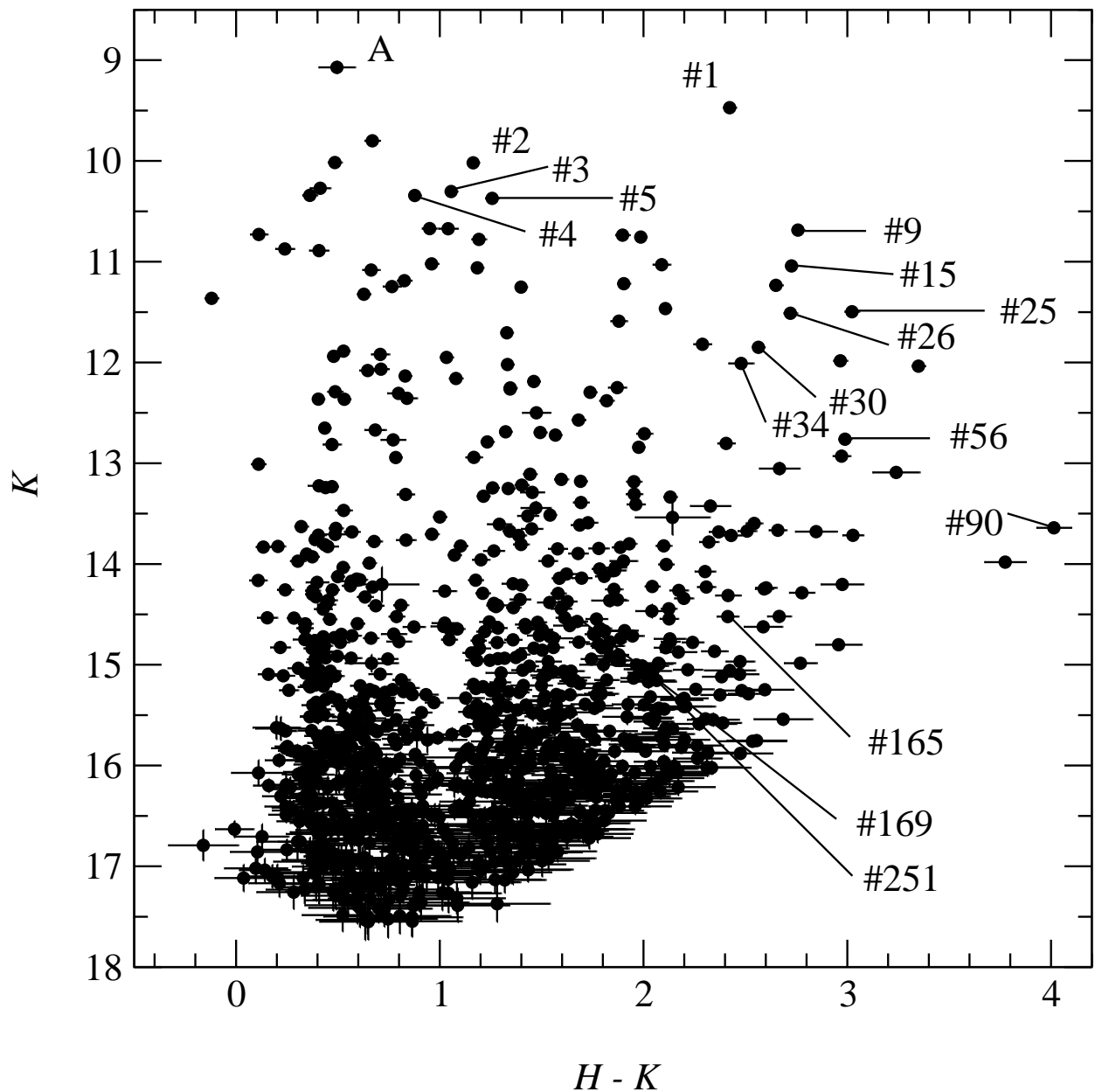} 
\figcaption[]{$H-K$
color--magnitude diagram (CMD) for the W31 cluster and surrounding
field. The stars labeled \#2 -- \#5 are O stars (see \S3.2.1). Stars
labeled \#1, 9, 15, 26, and 30 are massive young stellar object
candidates; see text and Figure~\ref{cc}.  The remaining labeled
objects are candidate counterparts to 5 GHz radio sources; see text.
\label{cmd}
}
\end{figure}

\begin{figure}
\figurenum{4} 
\plotone{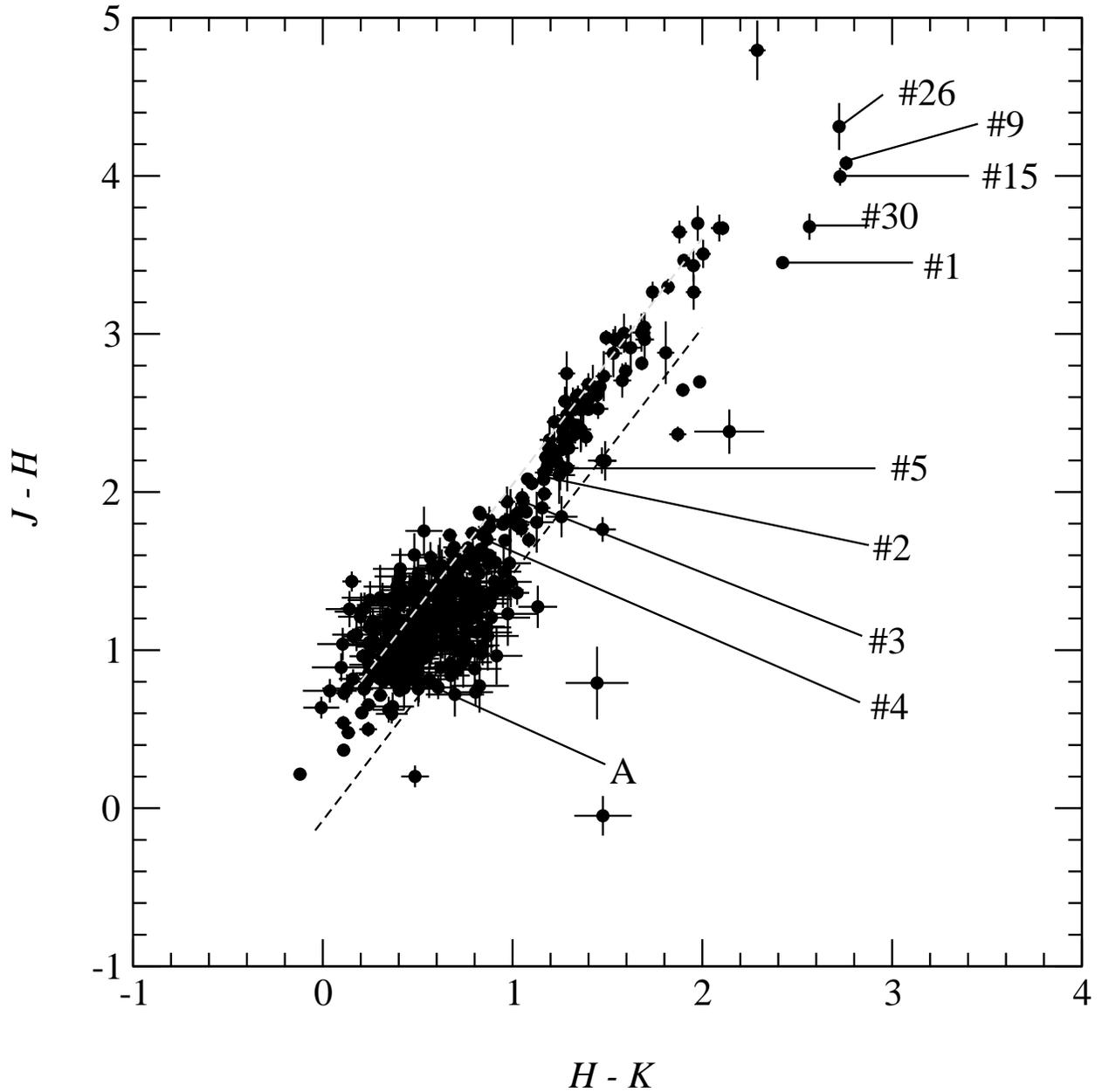} 
\figcaption[]{$J-H$ vs. $H-K$
color--color diagram for the the W31 cluster. The two {\it dashed}
lines represent the reddening lines for O and M stars using the
interstellar extinction law of \citet{m90}. See Figure~\ref{cmd}
for an explanation of labeled objects. An expected color--correction to
the OSIRIS photometry would bring the sequence in which these stars
lie down toward the lower reddening line; see text.
\label{cc}
}
\end{figure}

\begin{figure}
\figurenum{5} 
\plotone{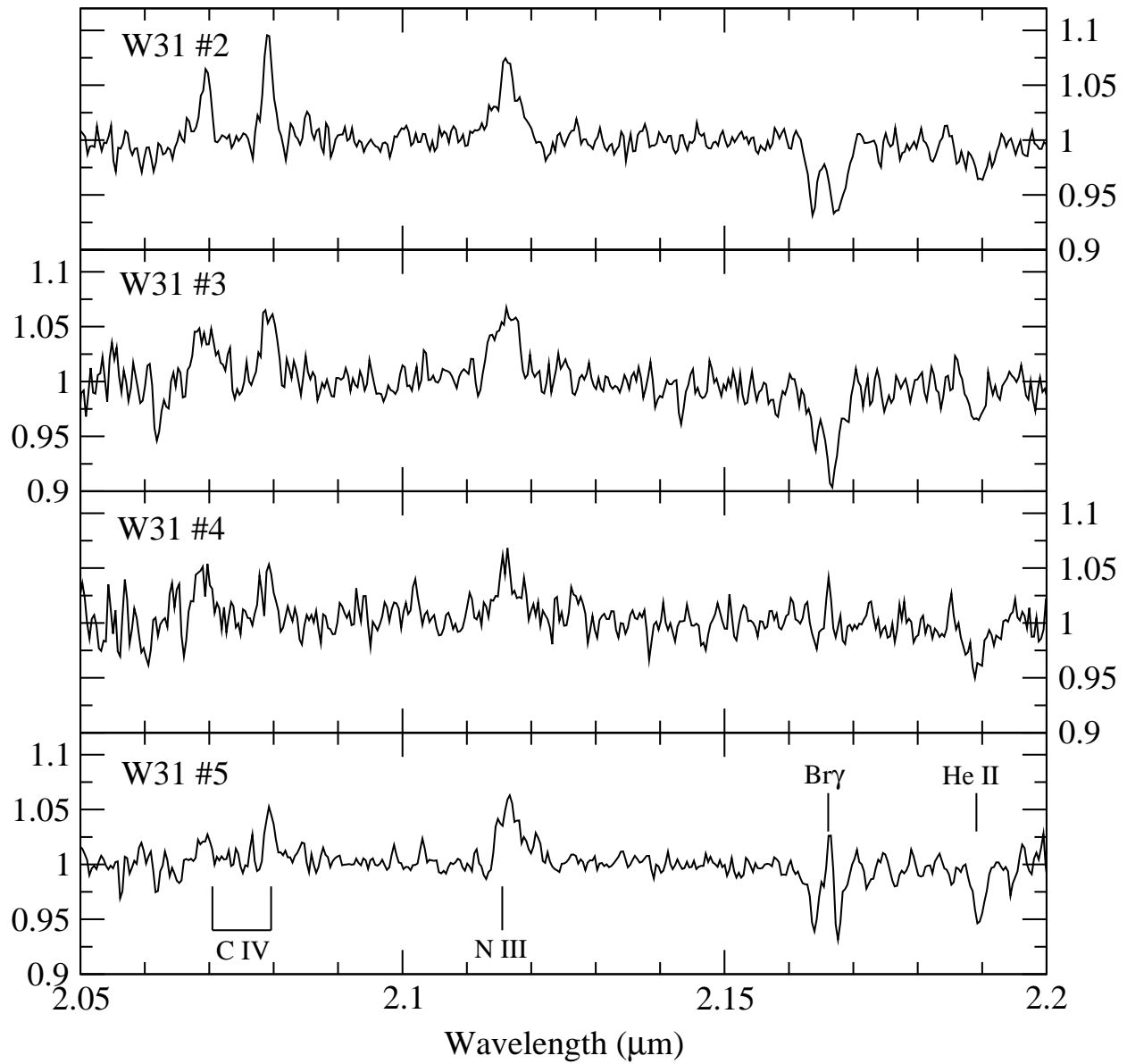} 
\figcaption[]{ $K-$band spectra
for the brightest stars in the W31 cluster.  The two pixel resolution
is $\lambda/\Delta\lambda \approx$ 3000. The spectra were summed in
apertures 0.64$''$ wide $\times$ a slit width of 0.48$''$ and include
background subtraction from apertures centered \aple 1.0$''$ on either
side of the object.  Each spectrum has been divided by a low order fit
to the continuum (after correction for telluric absorption; see \S2).
\label{spec}
}
\end{figure}

\begin{figure}
\figurenum{6} 
\plotone{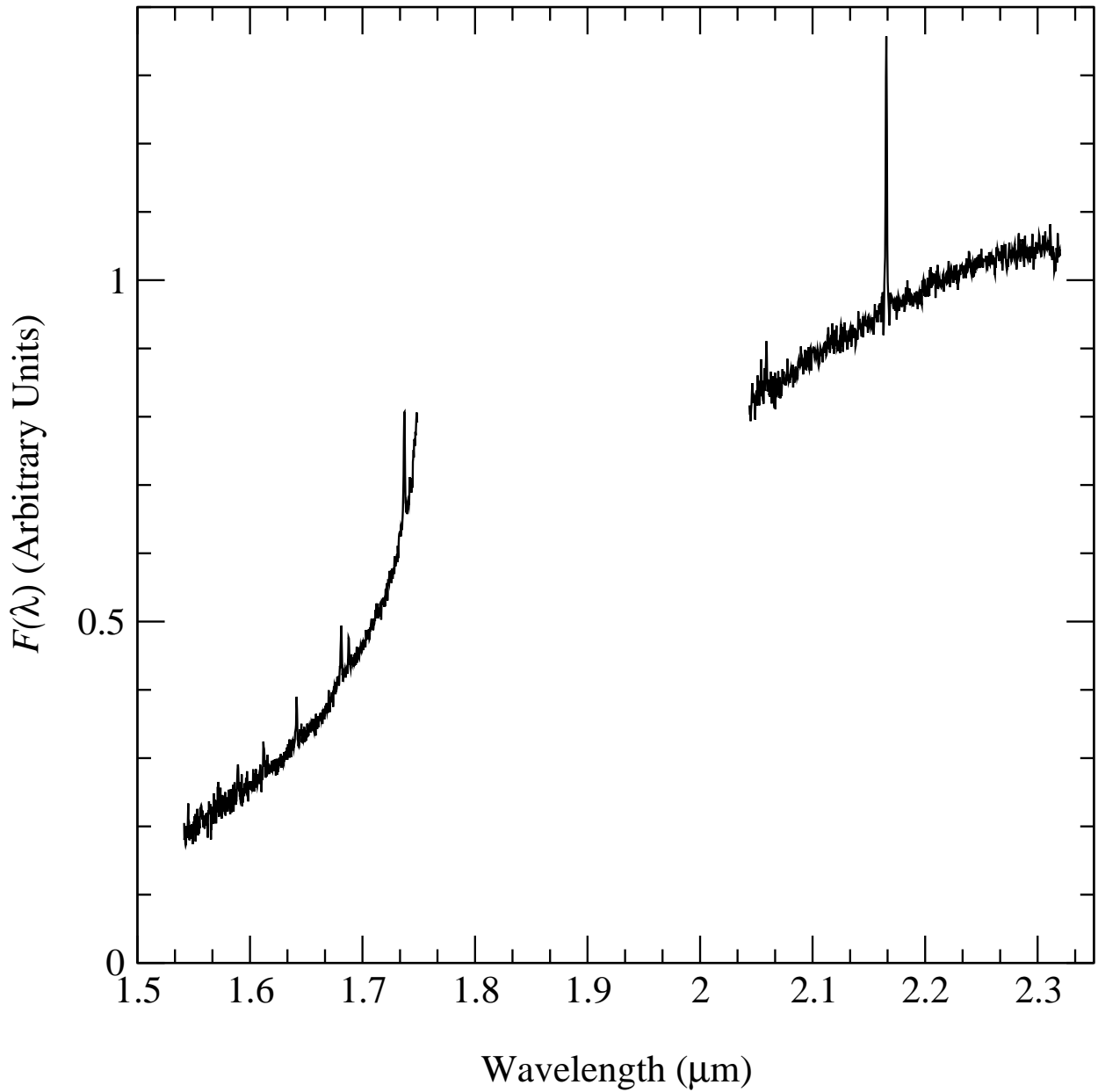} 
\figcaption[]{$H$ and $K-$band
spectra for the brightest star in the W31 cluster: W31~\#1 (excluding
foreground stars; see text).  The two pixel resolution is
$\lambda/\Delta\lambda \approx$ 3000. The spectra were summed in
apertures 0.64$''$ wide $\times$ a slit width of 0.48$''$ and include
background subtraction from apertures centered \aple 1.0$''$ on either
side of the object.  
The spectra are not corrected for extinction, and have been
adjusted in relative intensity using the
derived photometry ($H$ and $K$ magnitudes) for W31~\#1.
\label{hspec}
}
\end{figure}

\begin{figure}
\figurenum{7} 
\plotone{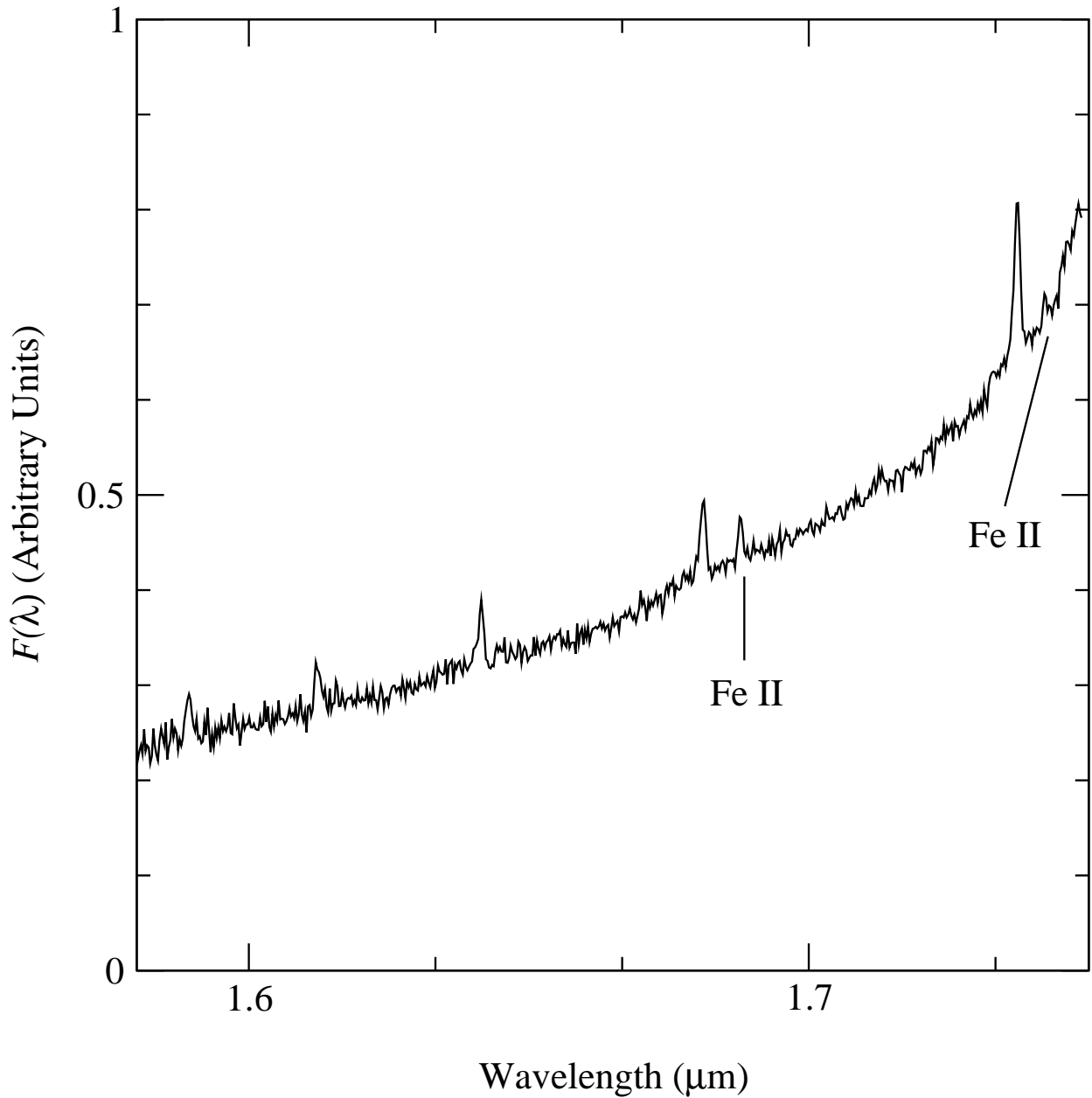} 
\figcaption[]{Blow--up of the
$H-$band spectrum shown in Figure~\ref{hspec} for W31~\#1. An
\ion{Fe}{2} line is present at 1.6878 \mic \ and perhaps also at 1.7419 
\mic. Brackett series lines of \ion{H}{1} at 1.7379 \mic, 1.6811 
\mic, 1.6412 \mic, 1.6113 \mic \ and 1.5885 \mic \ are
present.
\label{feii}
}
\end{figure}

\begin{figure}
\figurenum{8} 
\plotone{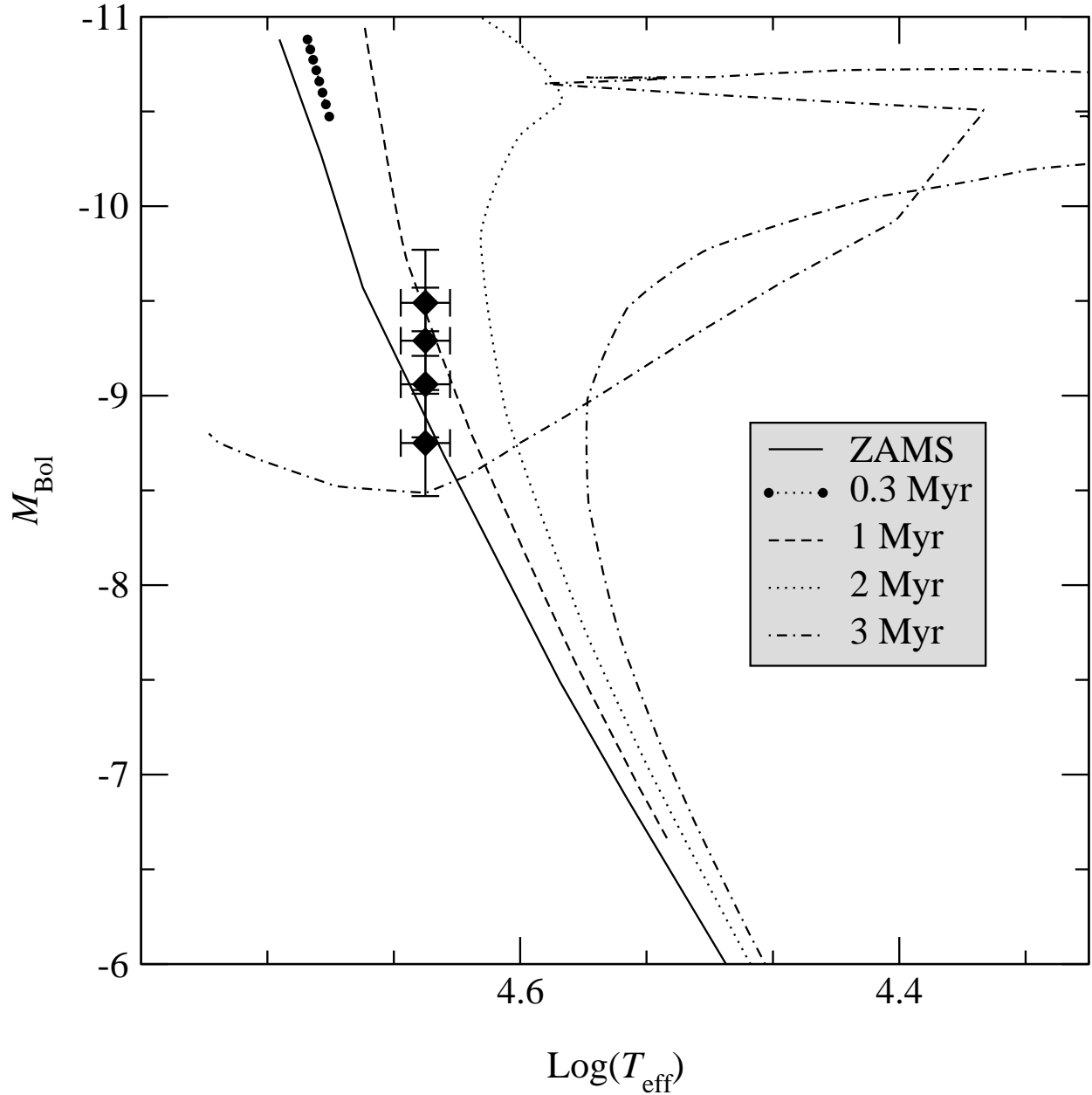} 
\figcaption[]{Hertzprung--Russell
diagram for the O stars in W31. The \teff \ is derived using the
spectral type vs. \teff \ relation given by \citet{vgs96}. $M_{\rm
bol}$ is derived using the distance estimate of \S4, the bolometric
correction given by \cite{vgs96}, the photometry of this paper, and
the $V - K$ tabulated by \citet{k83}. The isochrones are taken from
the models of \citet{ssmm92}.
\label{hrd}
}
\end{figure}

\begin{figure}
\figurenum{9} 
%\plotone{Blum.fig9.eps} 
\figcaption[]{Overlay of
Figure~\ref{w31a} $K-$band image with the \citet{geal89} 5 GHz radio
source positions (from their Table 2). See text and Table~\ref{ghosh}
for a discussion of individual sources and their near infrared
counterparts.  The $K-$band image center is $\alpha$~(2000) = 18$^{\rm
h}~09^{\rm m}~26.71^{\rm m}$, $\delta~(2000) = -20$\deg$~19'~29.7''$.
\label{radio}
}
\end{figure}

\newpage
%TABLES
\pagestyle{empty}
\begin{deluxetable}{lrrrcc}
\tablecaption{O Star Properties\label{ostar}} \tablewidth{0pt}
\tablehead{ \colhead{ID} & 
\colhead{$K$ \tablenotemark{a} } &
\colhead{$H - K$ \tablenotemark{a} } & 
\colhead{$A_K$ \tablenotemark{b} } & 
\colhead{$D_{ZAMS}(kpc)$ \tablenotemark{c}} &
\colhead{$D_V$}(kpc) \tablenotemark{c} } 
\startdata
W31~\#2&10.02$\pm$0.02&1.16$\pm$0.03&1.82$\pm$0.16&2.6&3.1 \\
W31~\#3&10.30$\pm$0.02&1.06$\pm$0.03&1.67$\pm$0.16&3.2&3.8 \\
W31~\#4&10.34$\pm$0.02&0.88$\pm$0.03&1.40$\pm$0.16&3.7&4.4 \\
W31~\#5&10.37$\pm$0.03&1.26$\pm$0.04&1.97$\pm$0.16&2.9&3.4 \\
Average & & &1.71$\pm$0.19&3.1$\pm$0.3&3.7$\pm$0.4
\enddata \tablenotetext{a}{Uncertainty in photometry is the sum in
quadrature of the photometric uncertainty plus the PSF fitting
uncertainty; see \S2.}  \tablenotetext{b}{The uncertainty in $A_K$ is
dominated by the variation in the power--law exponent of the
\citet{m90} interstellar extinction law; see \S3.3. The uncertainty in
mean $A_K$ is the sum in quadrature of the standard deviation in the
mean plus the (0.15 mag) systematic uncertainty due to the extinction
law.}  \tablenotetext{c}{Distance estimates assuming mean ZAMS and dwarf
(V) luminosities; see text. The uncertainty in the
distance is taken as the sum in quadrature of the standard deviation
in the mean of the individual estimates plus a component ($\sim$
250--500 pc) due to the systematic uncertainty in $A_K$.}

\end{deluxetable}

\pagestyle{empty}
\begin{deluxetable}{lrrr}
\tablecaption{YSO Properties\label{yso}} \tablewidth{0pt}
\tablehead{ 
\colhead{ID} & 
\colhead{$K$ \tablenotemark{a} } &
\colhead{$H - K$ \tablenotemark{a} } &
\colhead{$J - H$ \tablenotemark{a} } } 
\startdata
W31~\#1&9.47 $\pm$0.03&2.42$\pm$0.03&3.45$\pm$0.04\\
W31~\#9&10.69$\pm$0.02&2.76$\pm$0.03&4.08$\pm$0.05\\
W31~\#15&11.04$\pm$0.02&2.73$\pm$0.03&4.00$\pm$0.06\\
W31~\#26&11.51$\pm$0.03&2.72$\pm$0.04&4.31$\pm$0.15\\
W31~\#30&11.85$\pm$0.02&2.56$\pm$0.03&3.68$\pm$0.08\\
\enddata 
\tablenotetext{a}{Uncertainty in photometry is the sum in
quadrature of the photometric uncertainty plus the PSF fitting
uncertainty; see \S2.}  
\end{deluxetable}

\pagestyle{empty}
\begin{deluxetable}{lrrrrrrr}
\tablecaption{Radio Source Infrared Properties\label{ghosh}} 
\tablewidth{0pt}
\rotate
\tablehead{ 
\colhead{ID} &
\colhead{Ghosh ID \tablenotemark{a} } & 
\colhead{$K$ \tablenotemark{b} } &
\colhead{$H - K$ \tablenotemark{b} } & 
\colhead{$J - H$ \tablenotemark{b} } &
\colhead{Offset ($''$) \tablenotemark{c}} &
\colhead{Log(Q$_{\circ}$) \tablenotemark{d} } &
\colhead{Radio SpTyp \tablenotemark{e}}
} 
\startdata
W31~\#1   & 23 & 9.47  $\pm$  0.03 & 2.42 $\pm$  0.03 & 3.45 $\pm$  0.04  
& 4.3&47.77&B0.5\\
W31~\#25  & 24 & 11.50 $\pm$  0.03 & 3.02 $\pm$  0.04 & \nodata& 
1.9&48.08&B0\\
W31~\#26  & 21 & 11.51 $\pm$  0.03 & 2.73 $\pm$  0.04 & 4.31 $\pm$  0.15  
& 3.0&48.29&O9.5\\
W31~\#34  & 25 & 12.01 $\pm$  0.05 & 2.48 $\pm$  0.06 & \nodata& 
5.3&47.72&B1\\
W31~\#56  & 16 & 12.76 $\pm$  0.02 & 2.99 $\pm$  0.03 & \nodata& 
5.6&48.11&B0\\
W31~\#90  & 19 & 13.64 $\pm$  0.02 & 4.01 $\pm$  0.09 & \nodata& 
3.6&47.50&\nodata\\
W31~\#165 & 17 & 14.52 $\pm$  0.03 & 2.41 $\pm$  0.06 & \nodata& 
3.9&47.84&B0.5\\
W31~\#169 & 22 & 14.57 $\pm$  0.04 & 1.67 $\pm$  0.05 & \nodata& 
4.3&48.00&B0\\
W31~\#251 & 14 & 15.08 $\pm$  0.05 & 1.99 $\pm$  0.07 & \nodata& 
2.7&48.07&B0\\
\nodata & 15 &\nodata &\nodata &\nodata &\nodata &\nodata &\nodata \\
\enddata 
\tablenotetext{a}{\citet{geal89} 5 GHz source ID (their Table 2)}
\tablenotetext{b}{Uncertainty in photometry is the sum in
quadrature of the photometric uncertainty plus the PSF fitting
uncertainty; see \S2.}
\tablenotetext{c}{Offset in arcseconds of the \citet{geal89} 5 GHz 
radio source from the nearest red infrared source of this paper}  
\tablenotetext{d}{Lyman continuum luminosity from \citet{geal89} but 
corrected
for our new distance of 3.4 kpc.}
\tablenotetext{e}{Spectral type corresponding to the Lyman continuum as 
derived
by \citet{vgs96}} 
\end{deluxetable}

\end{document}